\documentclass[aps,prb,twocolumn,nofootinbib,superscriptaddress]{revtex4-2}
\usepackage{amsfonts}
\usepackage{amscd}
\usepackage{amsmath}
\usepackage{amssymb}
\usepackage{txfonts}
\usepackage{here}
\usepackage[dvipdfmx]{graphicx}
\usepackage{dcolumn}
\usepackage{bm}
\usepackage{xcolor}
\usepackage[%
  colorlinks=true,
  urlcolor=blue,
  linkcolor=blue,
  citecolor=blue
]{hyperref}
\usepackage{tabularx}
\usepackage{etoolbox}
\usepackage{physics}
\usepackage[caption=false]{subfig}
\newcommand{\nn}{\nonumber}
\newcommand{\beq}{\begin{equation}}
\newcommand{\eeq}{\end{equation}}

\makeatletter
\let\cat@comma@active\@empty
\makeatother
\begin{document}

\title{
Manipulation of Majorana Kramers Qubit and its Tolerance in Time-Reversal-Invariant Topological Superconductor
}

\author{Yuki Tanaka}
\affiliation{Department of Materials Engineering Science, Osaka University, Toyonaka, Osaka 560-8531, Japan}
\author{Takumi Sanno}
\affiliation{Department of Materials Engineering Science, Osaka University, Toyonaka, Osaka 560-8531, Japan}
\author{Takeshi Mizushima}
\affiliation{Department of Materials Engineering Science, Osaka University, Toyonaka, Osaka 560-8531, Japan}
\author{Satoshi Fujimoto}
\affiliation{Department of Materials Engineering Science, Osaka University, Toyonaka, Osaka 560-8531, Japan}
\affiliation{Center for Quantum Information and Quantum Biology, Osaka University, Toyonaka, Osaka 560-8531, Japan}

\begin{abstract}
We investigate non-Abelian statistics of Majorana Kramers pairs (MKPs) in a network system of one-dimensional time-reversal invariant topological superconductors by using numerical simulations of braiding dynamics, and examine the tolerance against various perturbations which may cause decoherence of MKPs. We, first, consider effects of a magnetic field which breaks time-reversal symmetry. In contrast to a naive expectation, the non-Abelian braiding of 
MKPs is robust against applied magnetic fields provided that the initial and final states of a braiding process are invariant under the combination of a time reversal and a mirror reflection, even when intermediate states break the combined symmetry.
Secondly, we  investigate the stability of non-Abelian braidings in the case with gate-induced inhomogeneous potentials at junctions between superconducting nanowires, which generally generate a non-Majorana nearly zero-energy Andreeev bound state at the junctions.
It is found that the non-Majorana nearly zero-energy states interfere with MKPs, resulting in the failure of non-Abelian braidings, when the
length scale of the inhomogeneous potentials is comparable to the coherence length of the superconductors. 
\end{abstract}

\maketitle
\section{Introduction}
Majorana quasi-particles, which appear in a boundary of a topological superconductor, have attracted a lot of interest because of promising applications to fault-tolerant quantum computation~\cite{kitaevJSI321,Nayak2008,doi:10.1146/annurev-conmatphys-030212-184337,Alicea_2012,Leijnse_2012,satoJPSJ16,mizushimaJPSJ16,AasenPRX6}. 
A conventional $s$-wave superconductor with broken inversion symmetry and applied magnetic field is a promising 
candidate system of a topological superconductor which hosts Majorana particles~\cite{fujimotoPRB77,SatoPRL103,PhysRevLett.104.040502,Alicea_2010,PhysRevLett.105.077001,PhysRevLett.105.177002,StanescuIOP25}.
A system constructed by depositing a semiconductor on a superconductor
has been proposed as a typical platform for such systems, and
the experimental search for Majorana states based on this idea has been extensively attempted~\cite{Mourik_2012,Deng_2012,das12,PhysRevLett.119.136803,gul,deng16,finck,churchill,zhang2021large}.
Generally, a strong external magnetic field is required in realizing such topological superconductors with broken time-reversal symmetry. 
The necessity of a strong magnetic field may cause some difficulty of the stability of Majorana qubits, because a magnetic field reduces
quasiparticle energy gaps. 
This may lead to the vulnerability of the quantum states spanned by Majorana particles by quasi-particle poisoning in realistic situations~\cite{gol11,rai12,bud12,kar17,kna18,kna18v2,kar21}. 
In contrast to such time-reversal symmetry broken topological superconductors, there exist topological superconductors with time-reversal symmetry, which do not require an external magnetic field for realizing the topological phase~\cite{Wong_2012,Zhang_2013}. 
This type of topological superconductors can keep the topological gap close to the proximity-induced superconducting gap, which leads to longer decoherence time of the stored quantum information~\cite{Liu_2014}. 
However, researches on non-Abelian braiding dynamics in a realistic setup based on numerical simulations, in which quasi-particle poisoning and external causes of decoherence are taken into account, are still limited~\cite{Amorim_2015,Gao_2016,sekania17,xie20,andrzejPRB20,Sanno_2021,RainisPRB85,SchmidtPRB86,ChengPRB84,ScheurerPRB88,hon22,XuPRA78, XuPRA80}.

In this paper, we investigate the braiding dynamics of Majorana bound states in a time-reversal invariant topological superconductor (TRITSC),  which belongs to class DIII characterized by $\mathbb{Z}_2$ topological invariant~\cite{sch08,ShiozakiPRB90,HAIM20191}. 
We consider a topological superconductor nanowire with $d_{x^2-y^2}$ paring and the Rashba spin-orbit interaction~\cite{Wong_2012,SatoPRB82}. 
In this setup, Majorana quasi-particles cannot exist alone and must come in pairs, which are called Majorana Kramers Pairs (MKPs), due to the Kramers theorem. 
This point contrasts with the case of time-reversal symmetry broken topological superconductors classified as class D which require strong magnetic field for the realization of the topological phase. 
We particularly focus on the tolerance of non-Abelian braidings of MKPs against two types of perturbations which may cause decoherence of MKP quits. One is an applied magnetic field which breaks time-reversal symmetry, and may generate the energy gap of MKP states. 
The other one is a gate-induced inhomogeneous potential at junctions of superconducting nanowires, which may give rise to non-Majorana low-energy Andreev bound states~\cite{Kells_2012,liu17,Moo18,Moo18-2,pan20,pan20-2, ReegPRB98}.

The motivation of testing the former is related to the issue of how to read out MKP qubits. Because of the Kramers degeneracy,
the direct measurement of total fermion parity is not sufficient for reading out MKP qubits.
Although the detection scheme utilizing the phase-controlled Josephson effect is proposed before~\cite{Liu_2014}, technically simpler
methods are, if possible, desirable. A simple way of reading out MKP qubits is to apply a small magnetic field lifting the Kramers degeneracy, 
which enables us the detection of fermion parity of one partner of a MKP.
We examine the efficiency of this approach. 
With the use of dynamical simulations, it is found that, in contrast to a naive expectation, the non-Abelian braiding of 
MKPs is robust against applied magnetic fields provided that the initial and final states of a braiding process are invariant under the combination of a time reversal and a mirror reflection, even when intermediate states break the combined symmetry.

The effect of a gate-induced inhomogeneous potential, which is the second target of this paper, is crucially important for the detection
of Majorana zero-energy bound states in superconducting nanowire junction systems.
Recently, it has been pointed out that low-energy non-Majorana Andreev bound states can be induced by the inhomogeneous potentials at junctions,
and may give rise to nearly quantized conductance, which is believed to be a signature of Majorana bound states, 
even when the system is in a trivial phase~\cite{Kells_2012,liu17,Moo18,Moo18-2,pan20,pan20-2,zhang2021large}. Thus, it is necessary to discriminate between Majorana zero-energy states and trivial low-energy states by examining dependence of $I$-$V$ characteristic on controllable parameters. 
Also, in a topological phase, trivial low-energy states may interfere with Majorana zero-energy states, and disturb non-Abelian braidings.
We examine the tolerance of non-Abelian braidings against the interference effect.

The organization of this paper is as follows.
In Sec.~II, we demonstrate non-Abelian statistics of MKPs in an ideal superconductor nanowire junction system 
via numerical simulations of braiding dynamics.
In Sec.~III, we investigate the tolerance of non-Abelian statistics against applied magnetic fields and gate-induced inhomogeneous potentials.
Conclusion is given in Sec.~IV.


\section{BRAIDING OF MKPs}
\subsection{MKPs in $d_{x^2-y^2}$-wave superconductor}
\begin{figure}[t]
    \includegraphics[width=88mm]{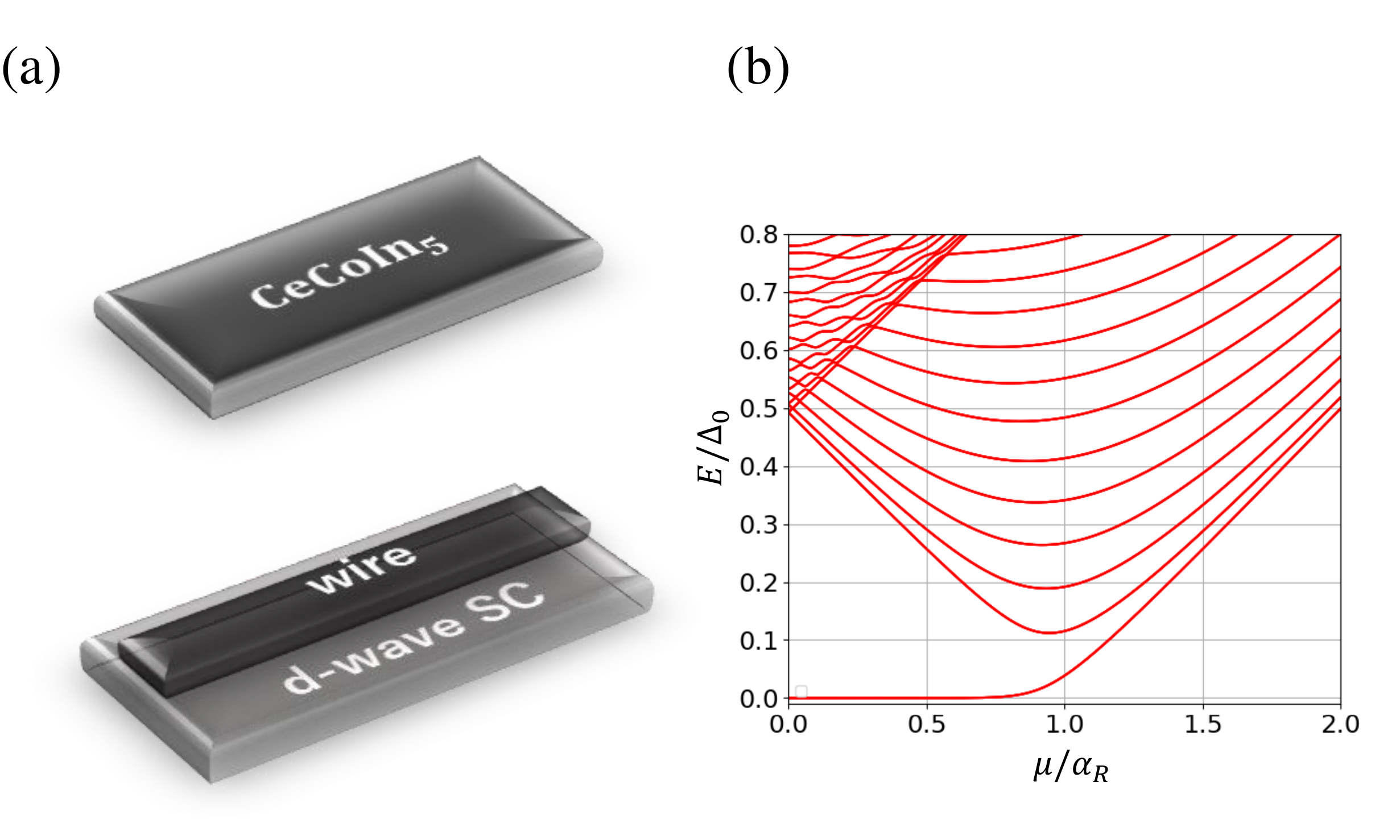}
    \includegraphics[width=88mm]{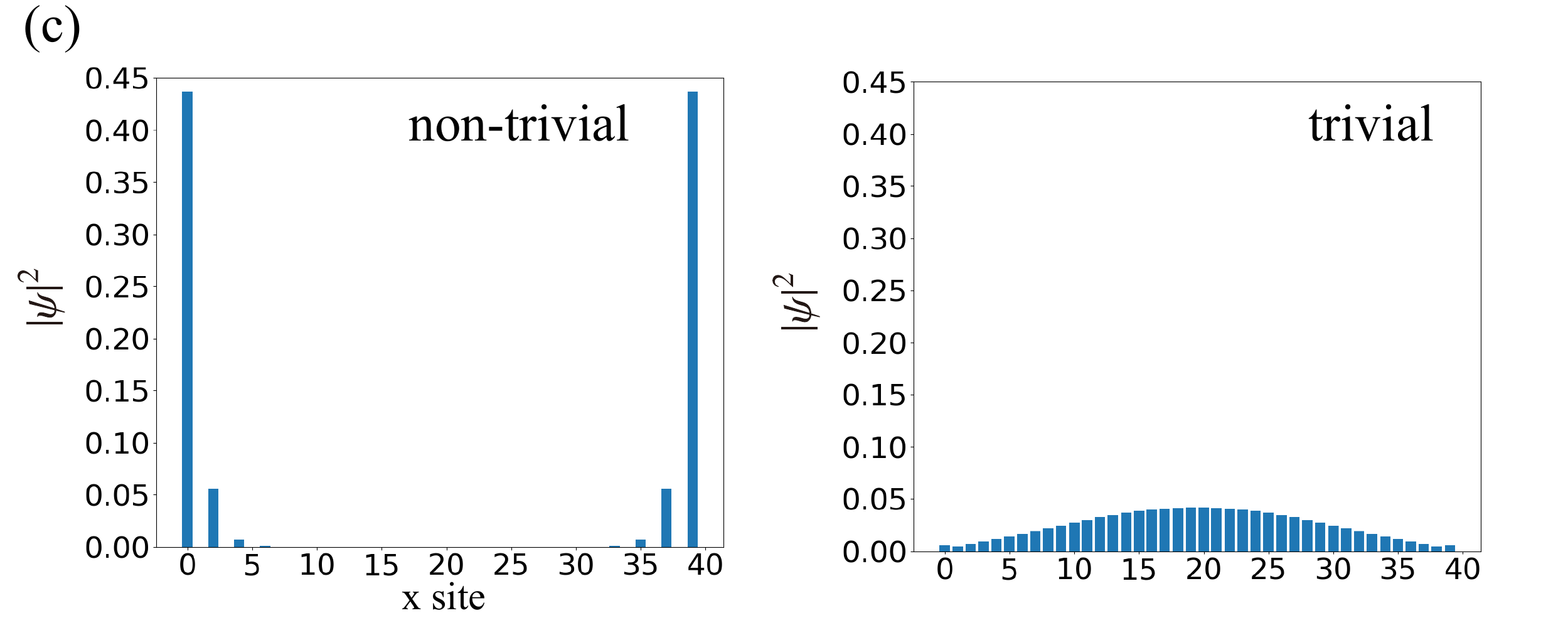}
    \caption{(a) Schematics of heterostructure systems which realize TRITSCs. Inversion symmetry breaking plays a crucial role for the TRITSCs. (b) The energy spectra of the single nanowire versus chemical potential $\mu$. Low-energy Majorana bound states exist for $\mu/\alpha_{\rm R}<1$. The parameters are set to be $t_{\rm h}=1$, $\alpha_{\rm R}=5$, $\Delta_0=10$ and $L_{\mathrm{wire}}=40$. Here, $L_{\mathrm{wire}}$ denotes the number of the sites of the single nanowire. (c) The total amplitude of the ground-state wave functions for topologically non-trivial and trivial cases. We set the chemical potential as $\mu=0.1$ and $\mu=7$ for each case. In the topologically non-trivial case, Majorana particles are located at the ends. In contrast, 
    there is no low-energy edge state for the trivial case. The low energy state in the bulk of the trivial case is due to the smallness of the bulk energy gap for $\mu=7$.}
    \label{fig:TRITSC}
\end{figure}

\begin{figure*}[t!]
    \centering
    \includegraphics[width=175mm]{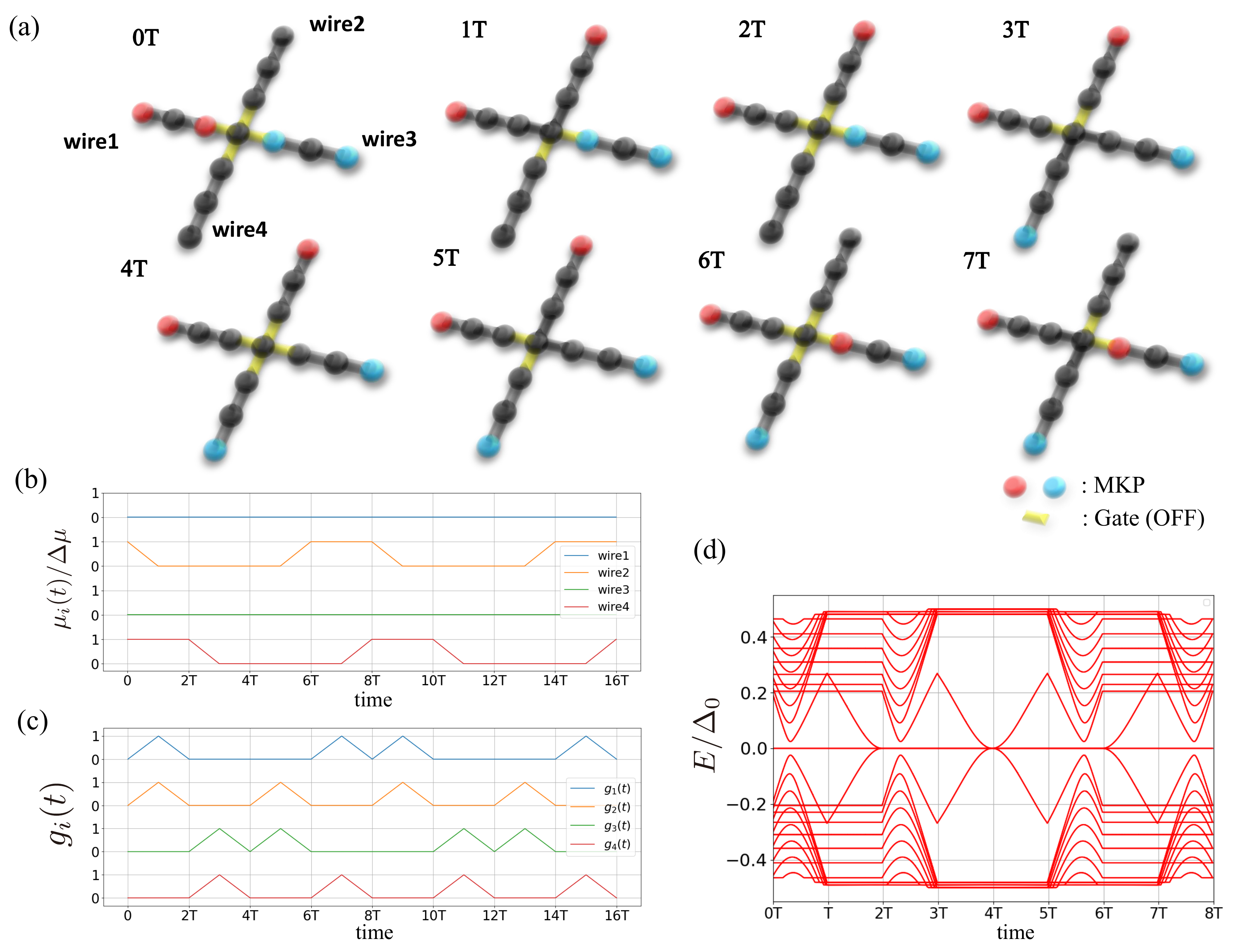}
    \caption{(a) Braiding process of two MKPs on the cruciform junction system. The red and blue circles represent MKPs which exist in the initial and final states of the wire 1 and wire 3, respectively. The MKPs that are generated in the intermediate steps of the braiding process are omitted in these figures.
    $8T$ is the period of the single braiding operation. (b)-(c) The instantaneous chemical potential and the gate potential of each wire. $\Delta \mathrm{cp}$ denotes the difference between the chemical potentials of the topological and trivial phases. Here, a wire is in the topological regime for $\mathrm{cp}/\Delta \mathrm{cp}\sim0$ while it is trivial if $\mathrm{cp}/\Delta \mathrm{cp}\sim1$. 
$\mathrm{gate}(t)=1$  corresponds to connecting a wire with the central site. (d) The quasiparticle energy spectra of the cruciform junction system in the braiding process (a). The instantaneous chemical potential and gate openness are set as shown in (b)-(c): $\mu_{\mathrm{min}}=0.1$; $\mu_{\mathrm{max}}=7$. The other parameters are $t_{\rm h}=1$, $\alpha_{\rm R}=5$, $\Delta_0=10$ and $L_{\mathrm{wire}}=40$.}
    \label{fig:braiding_and_parameter}
\end{figure*}

We consider one-dimensional TRITSC nanowire with proximity-induced $d_{x^2-y^2}$-wave pairings and the Rashba spin-orbit interaction, which hosts MKPs in the boundaries of the system. 
One of the candidates for a TRITSC is the surface of $d_{x^2-y^2}$-wave superconductors such as $\mathrm{CeCoIn_5}$~\cite{MovshovichPRL86,IzawaPRL87,KohoriPRB64,CurroPRB64,BianchiPRL89,SidorovPRL89,KakuyanagiPRL94,StockPRL100}, where inversion symmetry is broken~\cite{Wong_2012}. 
A conducting nanowire deposited on a noncentrosymmetric superconductor thin film is also a candidate system~\cite{Liu_2014,HAIM20191}. The schematics of these systems are illustrated in Fig.~\ref{fig:TRITSC}(a). 
In these nanowire systems, a phase boundary between the topological region and the trivial region can be moved via tuning chemical potential, which leads to the transfer of MKPs. Note that a single nanowire does not enable the exchange of two MKPs because they collide with each other in the intermediate processes. 
We consider a cruciform junction system for simulating the braiding dynamics of MKPs, as shown in Fig.~\ref{fig:braiding_and_parameter}(a). This junction system consists of four gates and four nanowires, whose chemical potentials are tunable~\cite{bauer18}. The Hamiltonian describing the cruciform junction is given by,
\begin{align}
\label{eq:hamiltonian}
&H = H_{\rm t} + H_{\rm SOC} + H_{\rm SC}, \\
&H_{\rm t} = -t_{\rm h} \sum_{{\bm{R}, \bm{d}},i,\sigma}  \left(
\psi^{\dagger}_{\bm{R}+ \bm{d},\sigma} \psi_{\bm{R}, \sigma} + \mathrm{h.c.}\right) 
- \sum_{\bm{R},i,\sigma} \mu_i \psi^{\dagger}_{\bm{R}, \sigma}\psi_{\bm{R}, \sigma}, \label{eq:Ht} \\
&H_{\rm SOC} = \sum_{\bm{R}, \bm{d},\alpha, \beta}\left\{ 
-\frac{i}{2}\alpha_{\rm R} \psi^\dagger_{\bm{R} + \bm{d}, \alpha} 
\hat{\bm{z}} \cdot ({\bm \sigma}_{\alpha,\beta} \times \bm{d})\psi_{\bm{R}, \beta} +\mathrm{h.c.}\right\}, \label{eq:Hsoc} \\
&H_{\rm SC} = \frac{\Delta_0}{2} \sum_{\bm{R}} \left\{ 
(\psi^{\dagger}_{\bm{R}+ {\bm{d}_x},\uparrow} \psi^{\dagger}_{\bm{R}, \downarrow}
-\psi^{\dagger}_{\bm{R} + \bm{d}_x, \downarrow}\psi^{\dagger}_{\bm{R}, \uparrow}) \right. \nn \\ 
&\quad \quad \quad \quad  \left. - (\psi^{\dagger}_{\bm{R}+ {\bm{d}_y},\uparrow} \psi^{\dagger}_{\bm{R}, \downarrow}
-\psi^{\dagger}_{\bm{R} + \bm{d}_y, \downarrow}\psi^{\dagger}_{\bm{R}, \uparrow})
+\mathrm{h.c.}\right\},
\end{align}
where $t_{\rm h}$, $\mu_i$, $\alpha_{\rm R}$ and $\Delta_0$ are the hopping integral, the chemical potential on wire $i$ ($i=1,2,3,4$), the spin-orbit coupling strength and the superconducting paring amplitude, respectively, and $\hat{\bm{z}}=(0,0,1)$.
Here, the subscript $\alpha$ and $\beta$ are the spin indices $\psi^{\dagger}_{\bm{R}, \alpha}(\psi_{\bm{R}, \alpha})$ is a creation (an annihilation) operator of an electron with spin $\alpha$ at $\bm{R}$,
i.e. the coordinates on the cruciform junction system.
${\bm \sigma}=(\sigma_x, \sigma_y, \sigma_z)$ is the Pauli matrix in the spin space. 
$\bm{d}$ denotes the two unit vectors $\bm{d}_x$ and $\bm{d}_y$, which are along the $x$ and $y$ directions. 
The $d$-wave gap function of this model can be represented as $\Delta({\bm k})=\Delta_0(\cos k_x -\cos k_y)$ in the momentum space. In addition, to simulate
dynamics of braiding processes,
we introduce time-dependent parameters $g_i(t) \in [0,1]$ $(i = 1,2,3,4)$, which represent the openness/closeness degree of each gate at the junctions between the central site of the cruciform system and the neighboring sites.
That is, at these junctions, $t_{\rm h}$, $\alpha_{\rm R}$, and $\Delta_0$ are multiplied by $g_i(t)$.
Each nanowire $i$ is in the topological phase for $|\mu_i|< \alpha_{\rm R}$, 
and in the trivial phase for $|\mu_i| > \alpha_{\rm R}$. In Fig.~\ref{fig:TRITSC}(b-c), we show the energy spectra of the single nanowire as a function of chemical potential $\mu$ and the total amplitude of the ground-state wave functions for topologically non-trivial and trivial cases. The low-energy modes for $\mu<\alpha_{\rm R}$ in Fig.~\ref{fig:TRITSC}(b) correspond to Majorana quasi-particle. It can be seen from Fig.~\ref{fig:TRITSC}(c) that Majorana quasi-particles are located at each end of the nanowire in the topologically non-trivial phase. 

To clarify the topology of MKPs, let us summarize the symmetries the Bogoliubov-de Gennes (BdG) Hamiltonian for the superconducting wire along the $x$-direction,
\begin{align}
{H}_{\rm BdG}(k_x) 
= \begin{pmatrix}
{h}_0(k_x) & i{\Delta}(k_x)\sigma_y \\
-i{\Delta}(k_x)\sigma_y & -h_{0}^{\rm tr}(-k_x)
\end{pmatrix},
\label{eq:bdg}
\end{align}
where ${h}_0(k_x)$ is the single-particle Hamiltonian density composed of Eqs.~\eqref{eq:Ht} and \eqref{eq:Hsoc} and we set $\mu _i \rightarrow \mu$. The Hamiltonian holds the particle-hole symmetry, ${\mathcal{C}} H_{\rm BdG}(k_x) {\mathcal{C}}^{-1} = -{H}_{\rm BdG}(-k_x)$, and the time-reversal symmetry, ${\mathcal{T}} H_{\rm BdG}(k_x) {\mathcal{T}}^{-1} = {H}_{\rm BdG}(-k_x)$,
where ${\mathcal{C}} = {\tau}_{x} K$ and $\mathcal{T}=-i\sigma_yK$ ($K$ is the complex conjugation operator) is the particle-hole operator, and $(\tau_x,\tau_y,\tau_z)$ are the Pauli matrices in the particle-hole space. Let $M_{xz}=i{\sigma}_{y}$ be the mirror reflection operator with respect to the $xz$-plane which flips the momentum and spin variables as $k_x\rightarrow k_{x}$, and ${\bm \sigma}\rightarrow (-\sigma_{x},\sigma_{y},-\sigma_{z})$. The BdG Hamiltonian is also invariant under mirror reflection symmetry,  
\begin{align}
{\mathcal{M}}_{xz}H_{\rm BdG}(k_x) {\mathcal{M}}^{ \dag}_{xz} = {H}_{\rm BdG}(k_{x}),
\label{eq:mirrorBdG}
\end{align}
where ${\mathcal{M}}_{xz} = {\rm diag}(M_{xz},M_{xz}^{\ast})$. Using these discrete symmetries, one can construct the chiral operator, ${\Gamma} =\mathcal{CTM}_{xz} =\tau_x$ and define the one-dimensional winding number as 
\begin{align}
w = - \frac{1}{4\pi i}\int^{+\pi/c}_{-\pi/c} d k_{x}{\rm tr}\left[ 
{\Gamma} H^{-1}_{\rm BdG}(k_x)\partial _{k_{x}}H_{\rm BdG}(k_x)
\right].
\label{eq:w1d}
\end{align} 
The topological invairiant is evaluated as  $w=-2$ for $\mu < \alpha_{\rm R}$, which counts the number of the zero energy states at the end of the superconducting wire~\cite{satoPRB2009,satoPRB11,miz12}. For $\mu<\alpha_{\rm R}$, the nontrivial value $|w|=2$ implies the existence of a pair of Majorana zero modes. We note that the chiral symmetry can be preserved unless the magnetic Zeeman field is applied perpendicular to the mirror plane, i.e., the $y$-axis.


\subsection{Numerical Methods}
The dynamics of braiding is described by the time-dependent BdG equation,
\begin{equation}
i\hbar \frac{\partial}{\partial t} \psi(t) = H_{\mathrm{BdG}}(t) \psi(t)
\end{equation}
where $\psi$ is the quasiparticle wave function in the Nambu representation.

We numerically solve this partial differential equation via the Chebyshev method~\cite{TalJSP81}. 
Generally, the time-propagation operator $\hat{U}(t+\Delta t; t)$ controlling the time-evolution of the system, which is defined by $\psi(t+\Delta t)=\hat{U}(t+\Delta t; t)\psi(t)$, can be written down as,
\begin{equation}
\hat{U}(t+\Delta t; t) = \hat{T} \exp \left [- i\int_t^{t+\Delta t} H_{\mathrm{BdG}}(\tau)d\tau \right].
\end{equation}
We note here that $\hat{U}(t+\Delta t; t)$ can be reduced to $\hat{U}(t+\Delta t; t) \approx \exp \left [-i H_{\mathrm{BdG}}(t) \Delta t \right ]$ for sufficiently small $\Delta t$. Under these conditions, we can expand the right-hand side with the Chebyshev polynomials as, 
\begin{equation}
\label{eq:Chebyshev}
\hat{U}(t+\Delta t; t) = \sum_{k=0}^{\infty} c_k(\Delta\tau)T_k(\tilde{H}(t)),
\end{equation}
where $\tilde{H}(t)$ and $\Delta\tau$ denotes $H_{\mathrm{BdG}}(t)/E_{\rm max}$ and $E_{\rm max}\Delta t$, respectively; $E_{\rm max}$ is the maximum eigenvalue of $H(0)$. This normalization is indispensable to avoid singularities of Chebyshev polynomials. The coefficient $c_k$ is defined as
\begin{align}
c_k(\Delta \tau)=
\left\{
\begin{array}{ll}
J_0(\Delta \tau) \quad &(k=0)\\
(-i)^k J_k(\Delta \tau) \quad &(k>0),
\end{array}
\right. , 
\end{align}
and $T_k$ is obtained from
\begin{gather}
T_{k+1}(\tilde{H}(t)) = 2\tilde{H}(t)T_k(\tilde{H}(t)-T_{k-1}(\tilde{H}(t)), \\
T_1(\tilde{H}(t))=\tilde{H}(t), \quad T_0(\tilde{H}(t))=\hat{1},
\end{gather}
where $J_k(x)$ is a $k$-degree Bessel function of the first kind. In numerical simulations, we can take the upper limit of the sum (\ref{eq:Chebyshev}) to be a finite value because of rapid decline in $c_k(\Delta \tau)$ with increasing $k$. This expansion provides the efficient approximation of the time-propagation operator.

For the simulation of braiding processes, we numerically demonstrate the twice exchange of MKPs at the inner end of the wire 1 and wire 3, which corresponds to a NOT-gate in the context of quantum computation. Fig.~\ref{fig:braiding_and_parameter}(a) illustrates the braiding process we simulate. 
The time-dependence of $\mu_i$ and $g_i(t)$ to realize this braiding process is shown in Figs.~\ref{fig:braiding_and_parameter}(b) and \ref{fig:braiding_and_parameter}(c). 
The instantaneous eigen energies in the braiding process are shown in Fig.~\ref{fig:braiding_and_parameter}(d). It can be seen that low-energy Majorana bound states exist under the specific conditions of the chemical potential $\mu$.

\begin{figure}[t!]
    \centering
    \includegraphics[width=90mm]{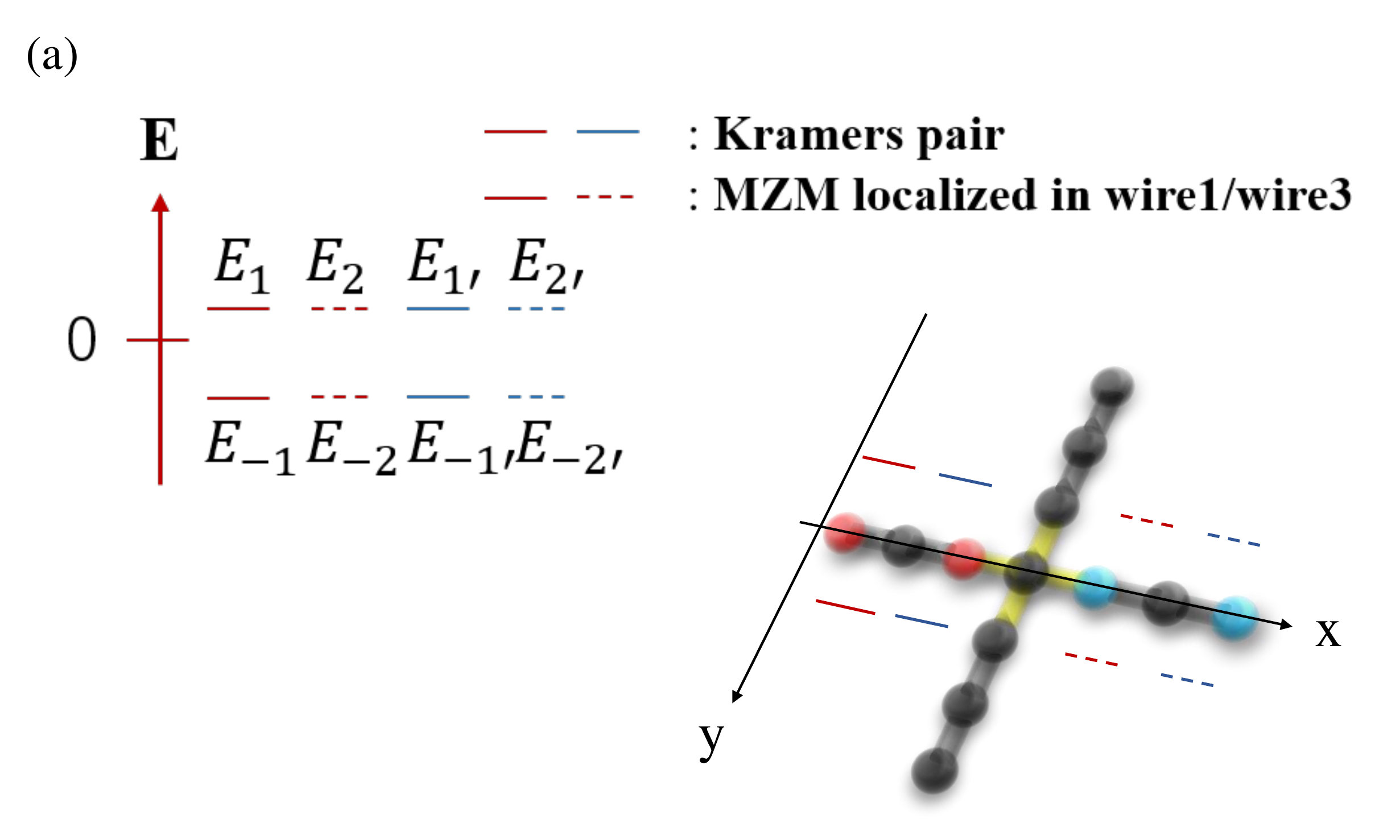}
    \includegraphics[width=90mm]{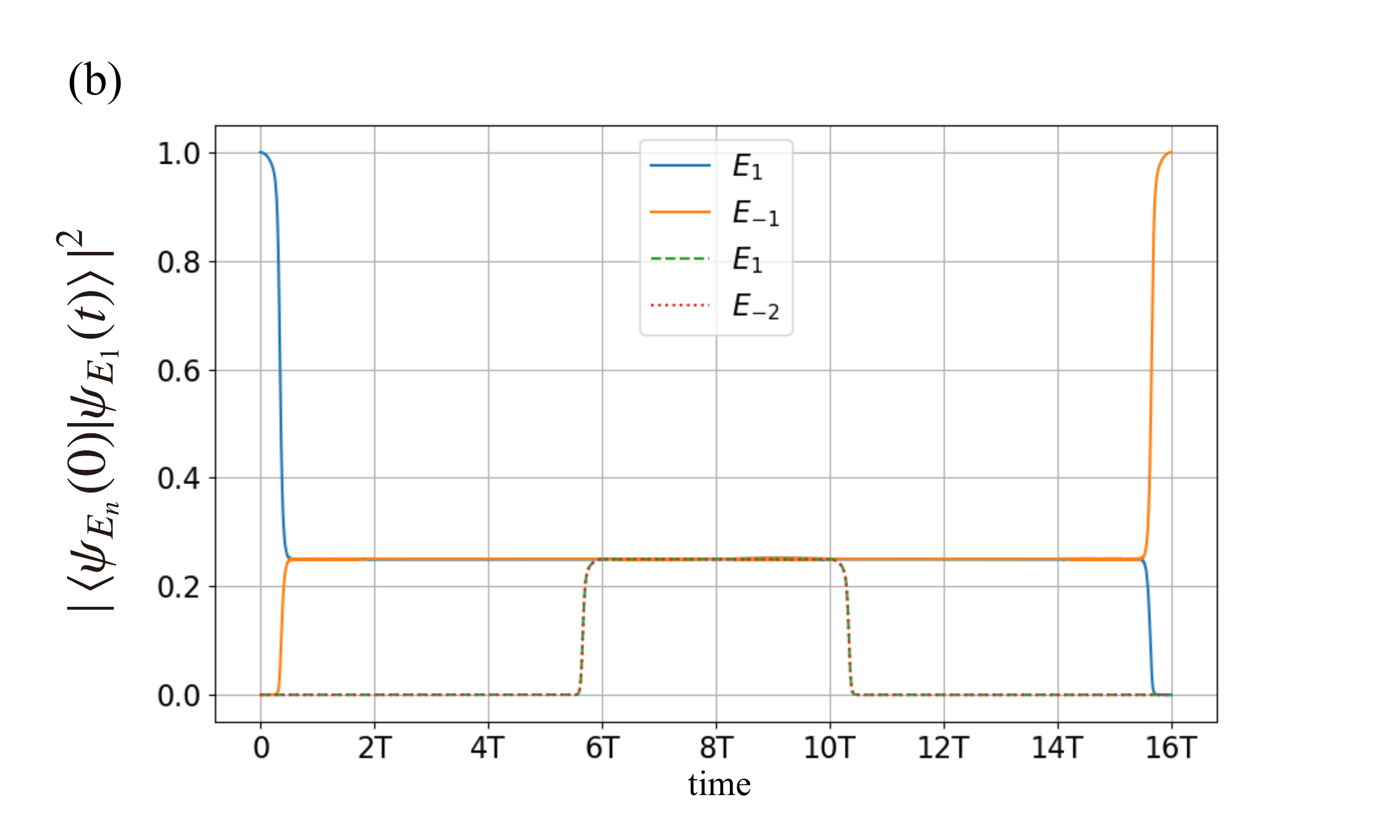}
    \includegraphics[width=90mm]{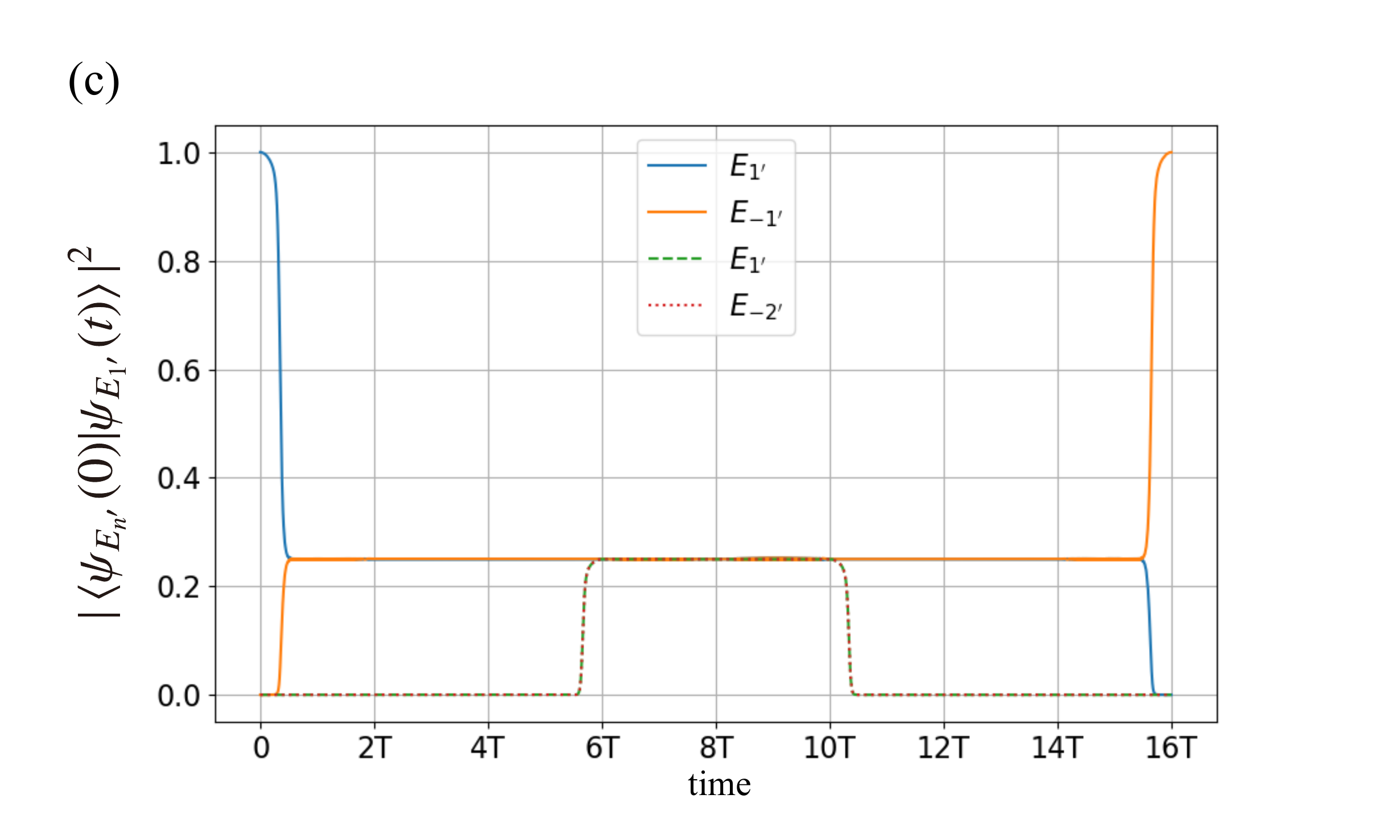}
    \caption{(a) The energy levels of MKPs in the initial state. 
    $E_1$ ($E_2$)and $E_{-1}$ ($E_{-2}$) are those of the wire 1 (wire 2), and $E_{1'}$ ($E_{2'}$) and $E_{-1'}$ ($E_{-2'}$) are their Kramers pair partners.
    (c) Projections of each initial eigenstate $\ket{\psi_{E_n}(0)}$ onto the instantaneous states $\ket{\psi_{E_{1'}}(t)}$. We set system parameters as $t_h=1$, $\mu_{\mathrm{min}}=0.1$, $\mu_{\mathrm{max}}=7$, $\alpha_R=5$, $\Delta_0=10$, $L_{\mathrm{wire}}=40$ and $T=500$.}
    \label{fig:label_and_fidelity}
\end{figure}


\subsection{Braiding MKPs in Superconducting Wires}
In the braiding protocol considered here, we switch between a topological phase and a trivial phase in the nanowires by tuning chemical potentials, which results in the transfer of Majorana quasi-particles~\cite{Alicea_2011}. This transfer method is based on the fact that Majorana quasi-particles exist at the open boundaries of a topological superconductor. The gate potential around the central site is tuned to connect and disconnect superconducting nanowires
with the central site.

For the simulations of the non-Abelian braiding corresponding to a NOT gate, we tune the chemical potentials of each wire $\mu_i$ and the gate parameters $g_i(t)$, as shown in Fig.~\ref{fig:braiding_and_parameter}(b-c). The wire $i$ is in a topological phase for $\mu_i/\Delta\mu\sim0$ while the wire is in a non-topological phase for $\mu_i/\Delta\mu\sim1$. $g_i(t)=0$ corresponds to turning off the connection between the wire $i$ and the central site. Here, $8T$ is the period of the single braiding operation. Generally, the twice exchange of two of four Majorana particles results in a NOT gate. In our setup, this is realized by the {\it twice} exchange of Majorana quasi-particles in the center side on the wire 1 and wire 3 in the initial states. The single exchange is done as follows. Initially, 
the wire 1 and wire 3 are in the topological phase, while the wire 2 and wire 4 are in a trivial phase. Then,
the Majorana quasi-particle located at the right end of the wire 1 is transported to the top end of the wire 2 by making the wire 2 topological via the chemical potential tuning, and gradually opening the gate 1 and gate 2 ($t=0\sim T$). After this operation is completed, the gates are gradually shut down ($t=T\sim2T$). Next, the Majorana quasi-particle on the wire 3 is moved to the bottom end of the wire 4 in the same way ($t=2T\sim4T$). In the next step, connecting the wire 2 and wire 3 via opening the gate 2 and gate 3 gradually  ($t=4T\sim5T$), the Majorana quasi-particle is moved to the wire 3 by changing the wire 2 into a trivial phase via chemical potential tuning ($t=5T\sim6T$).  In a similar manner, the Majorana quasi-particle on the wire 4 is transported to the wire 1 ($t=6T\sim7T$). Finally, the system returns to the same situation as the initial state with  the gate 1 and gate 4 gradually being close ($t=7T\sim8T$). These steps constitute the {\it single} braiding of Majorana quasi-particles.

\begin{figure*}[t!]
    \centering
    \includegraphics[height=150mm]{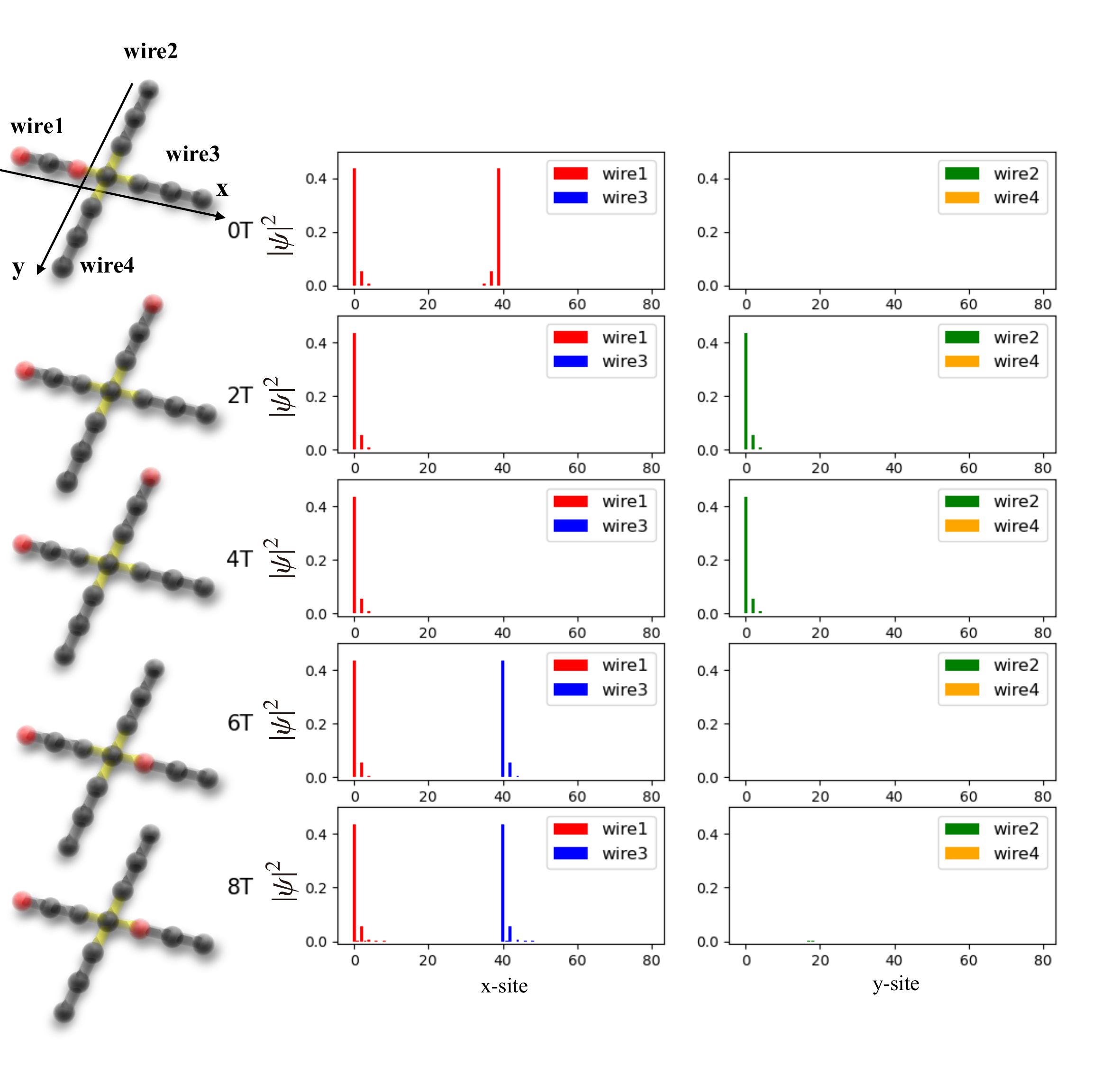}
    \caption{Time evolution of the profile of the wave function of low-energy bound states for a successful non-Abelian braiding.
    The braiding process of the state belonging to one side of the time-reversal partners of MKPs in the  wire 1 is shown. The vertical axis of each graph denotes the amplitude of the wave function. The graphs capture the moments of $0T$, $2T$, $4T$, $6T$ and $8T$. We take system parameters as $t_h=1$, $\mu_{\mathrm{min}}=0.1$, $\mu_{\mathrm{max}}=7$, $\alpha_R=5$ and $\Delta_0=10$. The spatial width of the Majorana end-state is $\xi\sim1$, which isd nearly equal to the superconducting coherence length.} 
    \label{fig:braiding_process_success}
\end{figure*}

\subsection{Results for an ideal case}
The numerical results for the braiding process of a NOT gate are shown in Fig. \ref{fig:label_and_fidelity}.
The projections of each initial eigenstate $\ket{\psi_{E_n}(0)}$ onto the instantaneous states $\ket{\psi_{E_1}(t)}$, which is given by time-evolution of the initial eigenstate $\ket{\psi_{E_1}(0)}$ are plotted in Fig.~\ref{fig:label_and_fidelity}(b). It can be seen that $\ket{\psi_{E_1}(0)}$ ends up not in the same state $\ket{\psi_{E_1}(0)}$ as the initial one, but changes into $\ket{\psi_{E_{-1}}(0)}$ after the braiding is completed ($t=16T$). A similar behavior is seen for the wave function $\ket{\psi_{E_{1'}}(t)}$ belonging to the other time-revered sector (Fig.~\ref{fig:label_and_fidelity}(c)). 
These numerical results verify that MKPs in our model indeed obey non-Abelian statistics. 
If we take the hopping integral as $t\sim0.1$ eV, the braiding operation period can be estimated as $T=100\sim10^{-12}$ [ps].

In addition, the phase shift predicted by the non-Abelian braiding rule of Majorana particles is confirmed as follows~\cite{Sanno_2021}:
\begin{align}
\mathrm{arg} \left[ \braket{\psi_{E_{-1}}(0)}{\psi_{E_1}(16T)} \right] &=-2.2\times 10^{-11}\\
\mathrm{arg} \left[ \braket{\psi_{E_{-2}}(0)}{\psi_{E_2}(16T)} \right] &=\pi - 6.9\times10^{-13}
\end{align}
Similarly, 
\begin{align}
\mathrm{arg} \left[ \braket{\psi_{E_{-1'}}(0)}{\psi_{E_{1'}}(16T)} \right]&=\pi -2.2\times 10^{-11}\\
\mathrm{arg} \left[ \braket{\psi_{E_{-2'}}(0)}{\psi_{E_{2'}}(16T)} \right]&= -6.9\times10^{-13}
\end{align}
The twice exchange of MKPs gives the wave function the phase shift $\pi$. 

We also show in Fig.~\ref{fig:braiding_process_success} the time evolution of the profile of the wave function of MZMs in the case of the successful braiding process corresponding to a NOT gate.

\section{Tolerance of Majorana Quantum Gates against Perturbations}

In this section, we investigate the tolerance of the quantum gates constructed from the braidings of MKPs.
We, particularly, focus on effects of applied magnetic fields which breaks time-reversal symmetry, and inhomogeneous gate potentials
which lead to non-Majorana low-energy bound states. These perturbations may cause decoherence of MKPs. 

\subsection{Effect of magnetic fields}
In order to investigate effects of magnetic fields on the braiding dynamics of MKPs, we introduce the Zeeman term $H_{q}$ into our Hamiltonian, which is given by
\begin{equation}
\label{eq:zeeman}
H_q = \sum_{\bm{R}, \alpha, \beta}V_q(\sigma_q)_{\alpha,\beta}\psi_{\bm{R},\alpha}^{\dagger}\psi_{\bm{R}, \beta}\  ,
\quad q =x,y,z, \forall t
\end{equation}
where $V_q$ is the strength of the Zeeman magnetic field, with $q$ the direction of the magnetic field and $\alpha$ and $\beta$ represent spin indices. 

\begin{figure}[t!]
    \centering
    \includegraphics[width=90mm]{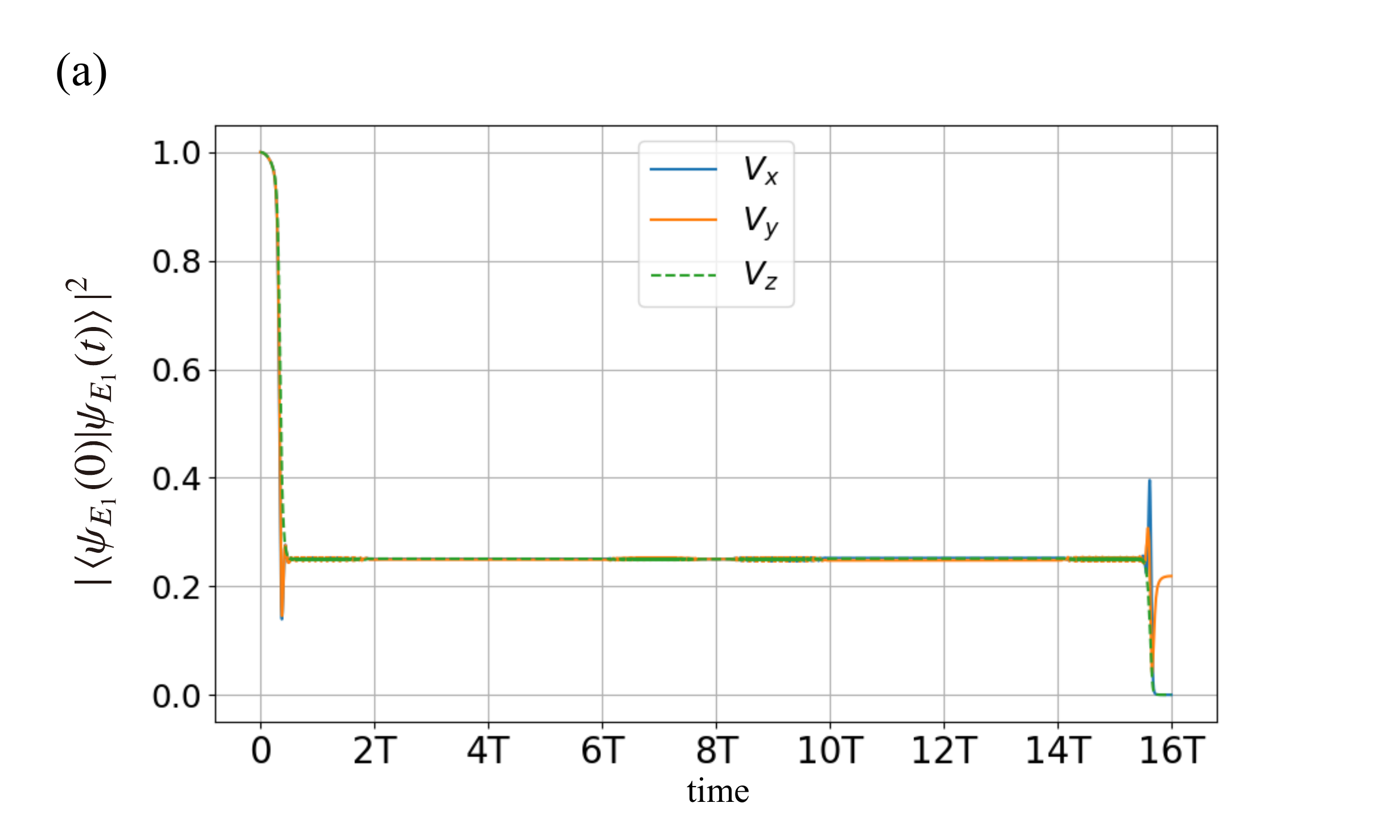}
    \includegraphics[width=90mm]{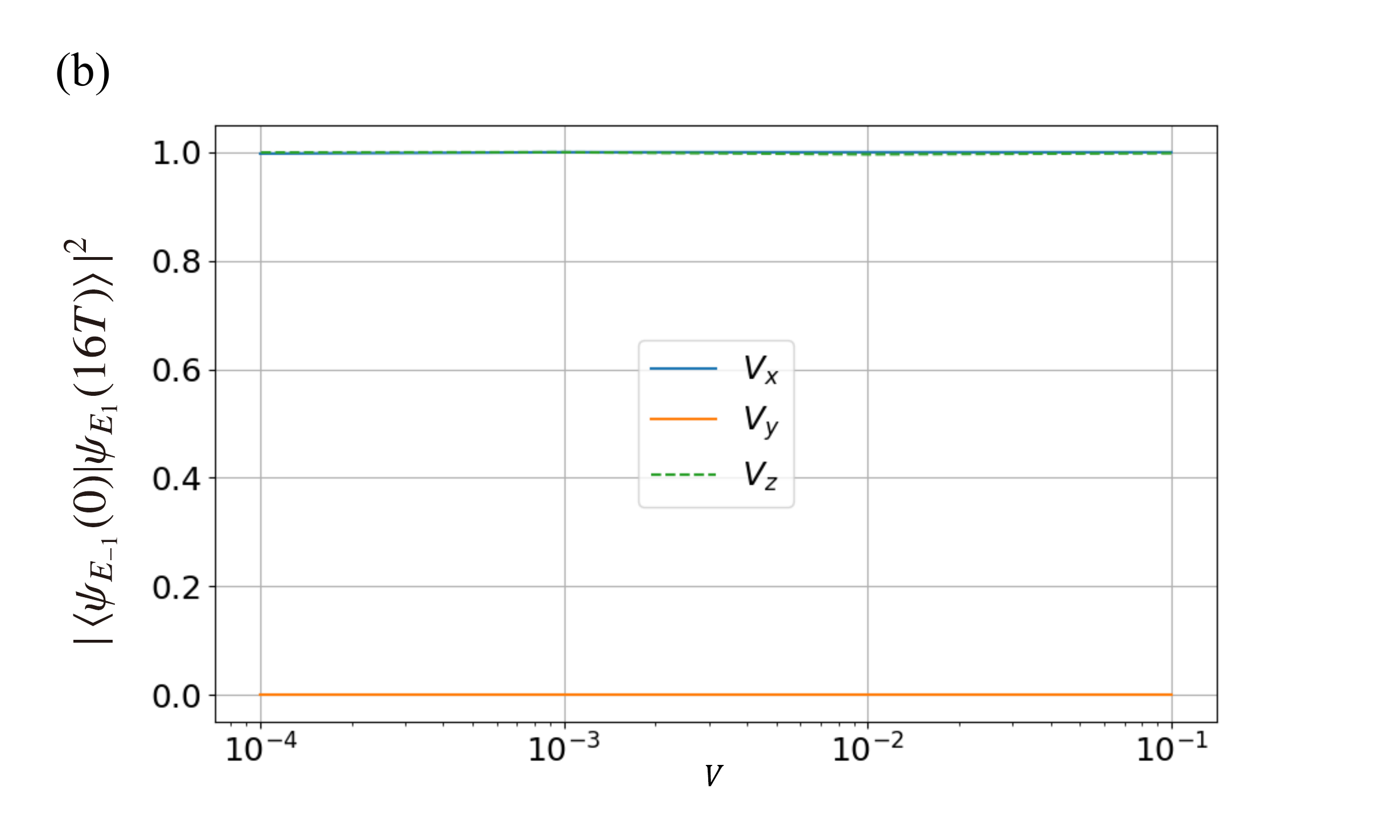}
    \caption{(a) Instantaneous fidelity $\abs{\braket{\psi_{E1}(0)}{\psi_{E1} (t) }}^2$ for $V_q/\Delta_0=0.01$. While the final fidelity $\abs{\braket{\psi_{E1} (0) }{\psi_{E1} (16T) }}^2$ should reduce to $0$ for the successful braiding, the final state in the case with a magnetic field along the $y$-direction results in a non-zero value, which implies the failure of the non-Abelian braidings.  (b) Transition probabilities plotted as a function of the strength of a magnetic field for magnetic fields parallel to the $x$, $y$ and $z$-directions.  For the magnetic field along the $x$ and $z$-directions, the non-Abelian braiding is successful, while for  the magnetic field along the $y$-direction, it is failed.
    We also note that the magnitude of the fidelity depends on the details of the system parameters when a magnetic field is applied along the $y$-direction. We set system parameters as $t_h=1$,  $\mu_{\mathrm{min}}=0.1$, $\mu_{\mathrm{max}}=7$, $\alpha_R=5$, $\Delta_0=10$, $L_{\mathrm{wire}}=40$ and $T=500$ for these calculations.}
    \label{fig:zeeman}
\end{figure}

As seen below, our results demonstrate that the direction of magnetic field is crucial for the braiding dynamics of MKPs. In Fig.~\ref{fig:zeeman}(a), we show the projection of the initial eigenstate $\ket{\psi_{E_1}(0)}$ onto the instantaneous state $\ket{\psi_{E_1}(t)}$, where $\ket{\psi_{E_1}(t)}$ is given by time evolution of $\ket{\psi_{E_1}(0)}$. We find nonzero values of projection $\abs{\braket{\psi_{E1} (0)}{ \psi_{E1} (16T) }}^2\neq 0$ after the manipulation completed when the magnetic field is applied along the $y$-direction. This implies that non-Abelian statistics of MKPs is failed because of breaking time-reversal symmetry. 
In contrast, magnetic fields applied along the $x$ and $z$-directions do not have a significant influence on the braiding dynamics, as shown in Fig.~\ref{fig:zeeman}(b). 
A key for understanding the origin of this unexpected result is the fact that 
the initial and final states of the braiding process are invariant under the combination of a time reversal and a mirror reflection when the magnetic field is applied along the $x$ and $z$-directions~\cite{miz13,tsu13,ShiozakiPRB90,miz15}.
As mentioned in Sec.~II, the nontrivial topology of the wires 1 and 3 is protected by the combined symmetry, which can be preserved unless the magnetic field is applied perpendicular to the mirror reflection ($xz$) plane. 
The nontrivial value of the winding number in Eq.~\eqref{eq:w1d}, $w=-2$, ensures the existence of MKPs even in the case with a magnetic field which breaks time-reversal symmetry. 
Note that in the intermediate states of the braiding protocol, the combined symmetry is not preserved. 

Interestingly, our numerical results shown in Fig.~\ref{fig:zeeman}(b) indicate that, in spite of the absence of symmetry protection in the intermediate states, the non-Abelian braiding is robust, and successful.
This is because the MKPs are not destroyed provided that the braiding period $8T$ is sufficiently smaller than the inverse of the Zeeman energy splitting of the MKPs.
This remarkable tolerance against magnetic fields may be utilized for reading out a Majorana qubit composed of only one Kramers pair partner by applying a magnetic field which preserves the combined symmetry.

\subsection{Gate-induced potential}
\begin{figure}[t!]
    \centering
    \includegraphics[width=90mm]{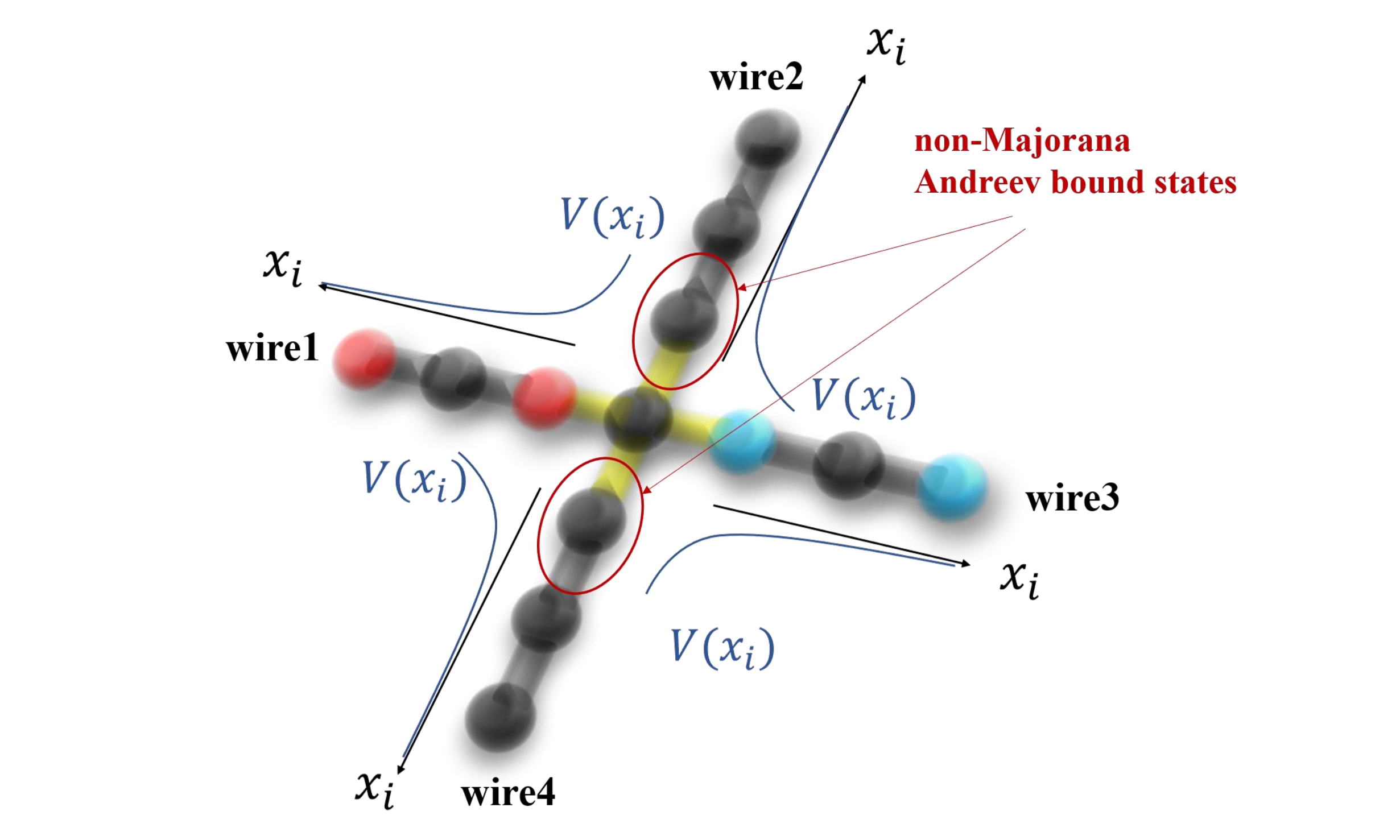}
    \caption{Schematic of the gate-induced potentials in our initial setup. We put the potentials in a radial direction from the central site. Non-Majorana Andreev bound states exist in the wire 2 and wire 4 
    in the initial states.}
    \label{fig:gate_induced_potentials}
\end{figure}

The smooth confinement potentials at junctions of superconducting nanowire systems can generally induce non-Majorana low-energy Andreev bound states~\cite{Kells_2012,liu17,Moo18,Moo18-2,pan20,pan20-2}.
It is expected that non-Majorana low-energy states interfere with MKPs, and 
prevent MKPs from properly moving in braiding processes, which leads to failure of the non-Abelian statistics. 
To examine this possibility, we introduce the following gate-induced inhomogeneous potentials to our Hamiltonian (\ref{eq:hamiltonian}). 
\begin{equation}
V(x_i) = V_0 \exp\left (- \frac{x_i ^2}{2\sigma^2} \right) \psi_{x_i}^{\dagger}\psi_{x_i}
\end{equation}
Here, the parameter $\sigma$ controls the characteristic length scale of spatial variation of the gate potentials and $x_i$ is the spatial coordinate of the $i$-th site with $x_0=0$ corresponding to the central site connecting the four nanowires. We put this potential in a radial direction from the central site, as illustrated in Fig.~\ref{fig:gate_induced_potentials}. 

\begin{figure}[t!]
    \includegraphics[width=90mm]{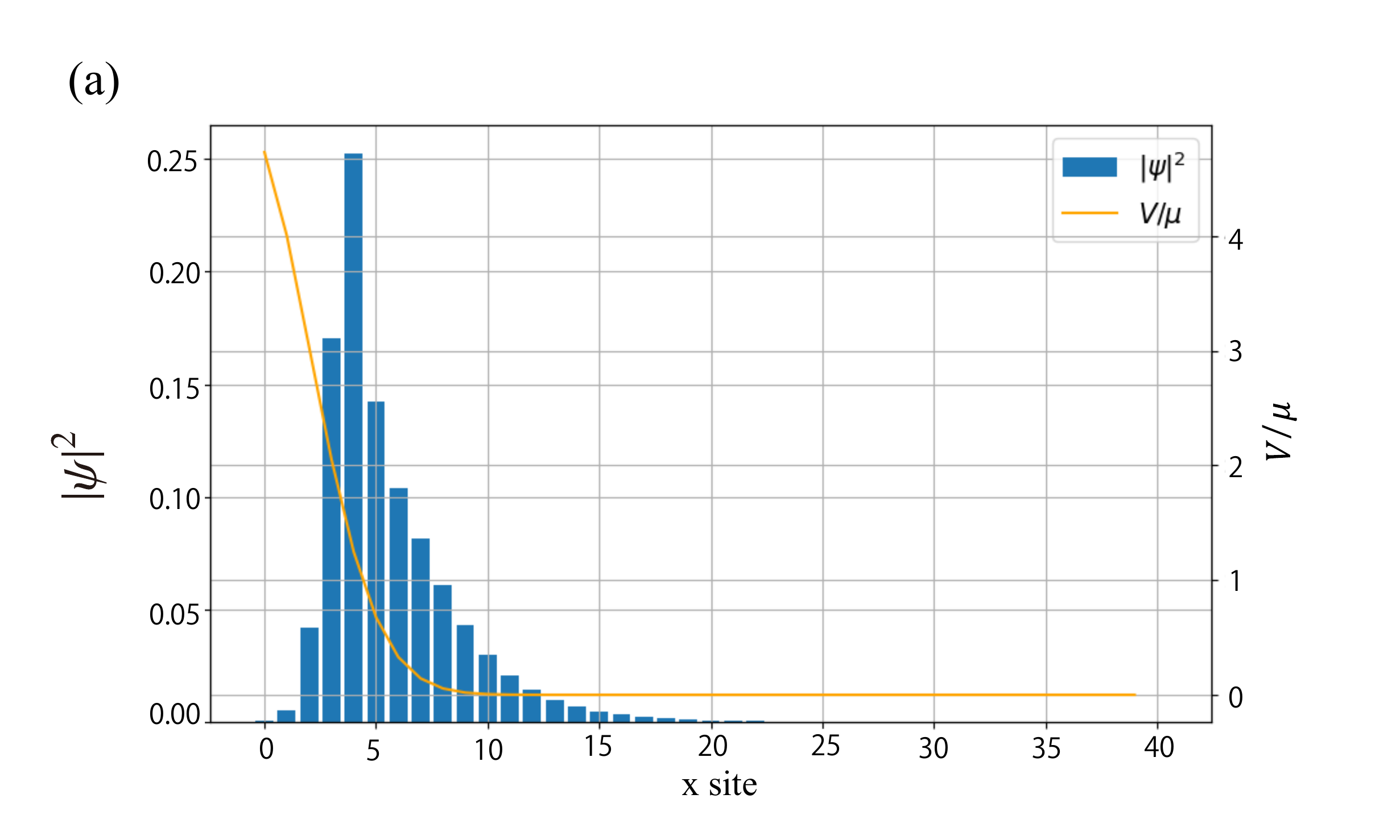}
    \includegraphics[width=90mm]{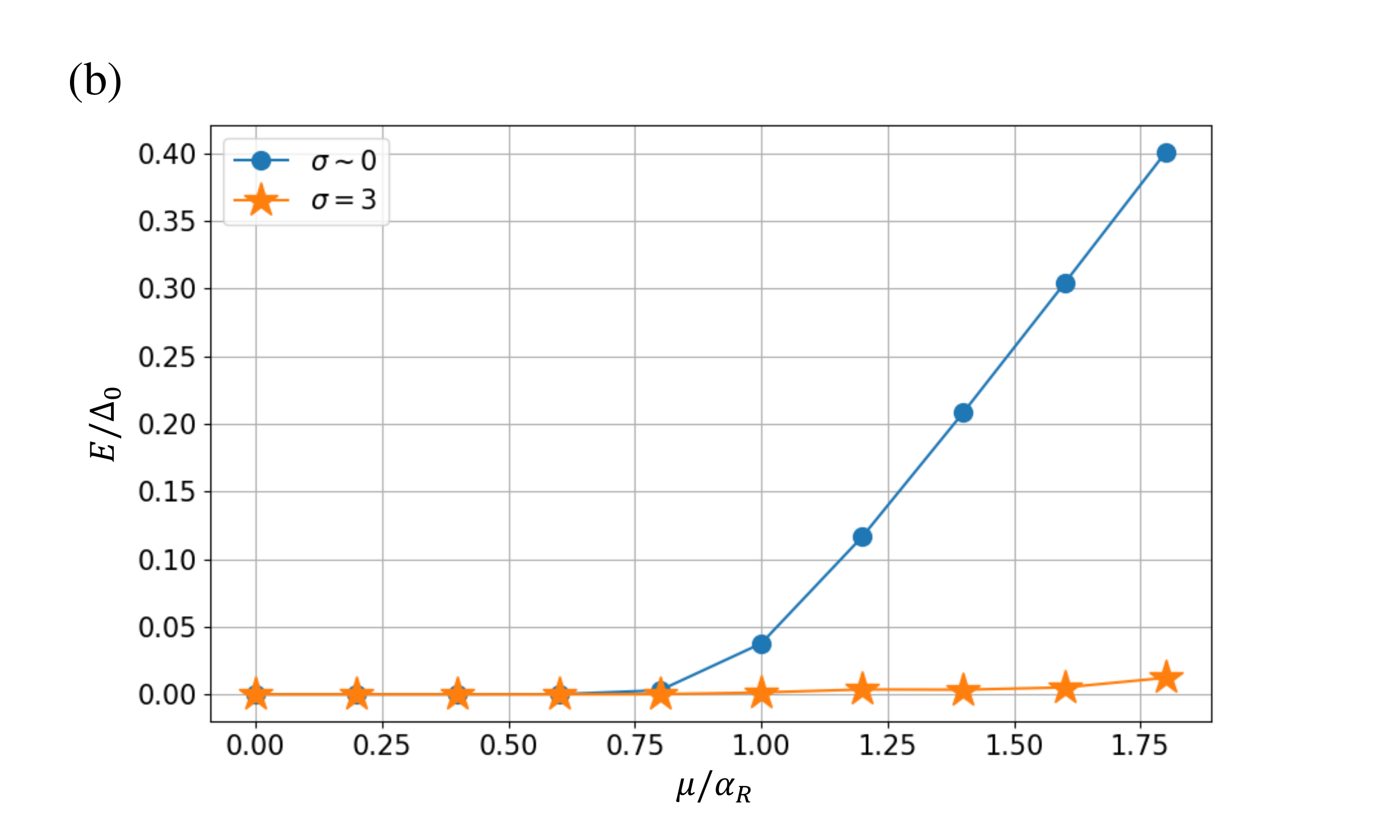}
        \caption{(a) The amplitude $|\psi|^2$ of non-Majorana Andreev Bound State and the confinement potential $V/\mu$ in a single wire of our setup
    versus the spatial coordinate $x$. 
    Note that this non-Majorana Andreev bound state exists in the wire 2 and the wire 4, even when they are in trivial phases.
    The parameters of this calculation are $t_h=1$, $\mu=7$, $\alpha_R=5$, $\Delta_0=10$, $L_{wire}=40$, $\sigma=3$ and $V_0=5\mu$. 
    (b) The Andreev end-state energy and Majorana end-state energy as a function of chemical potential $\mu$, for $\sigma\simeq0$ and $\sigma=3$. The parameters except $\mu$ and $\sigma$ are the same as (a). 
   It is seen that even in the non-topological regime $\alpha_R<\mu$, nearly zero-energy bound states which mimic Majorana bound states appear
    for $\sigma=3$, while they are pushed to higher-energy states for $\sigma\sim 0$.
  }
    \label{fig:abs_and_mzm}
\end{figure}

In Fig.~\ref{fig:abs_and_mzm}, we show
the profile of the wave function of the low-energy bound state caused by gate-induced potential and the energy level of the bound state.
  It is verified that the low-energy states exist in the vicinity of the gate in one-dimensional TRITSC nanowire even in the trivial phase with $\mu/\alpha_{\rm R}>1$. 

\begin{figure}[t!]
    \centering
    \includegraphics[width=90mm]{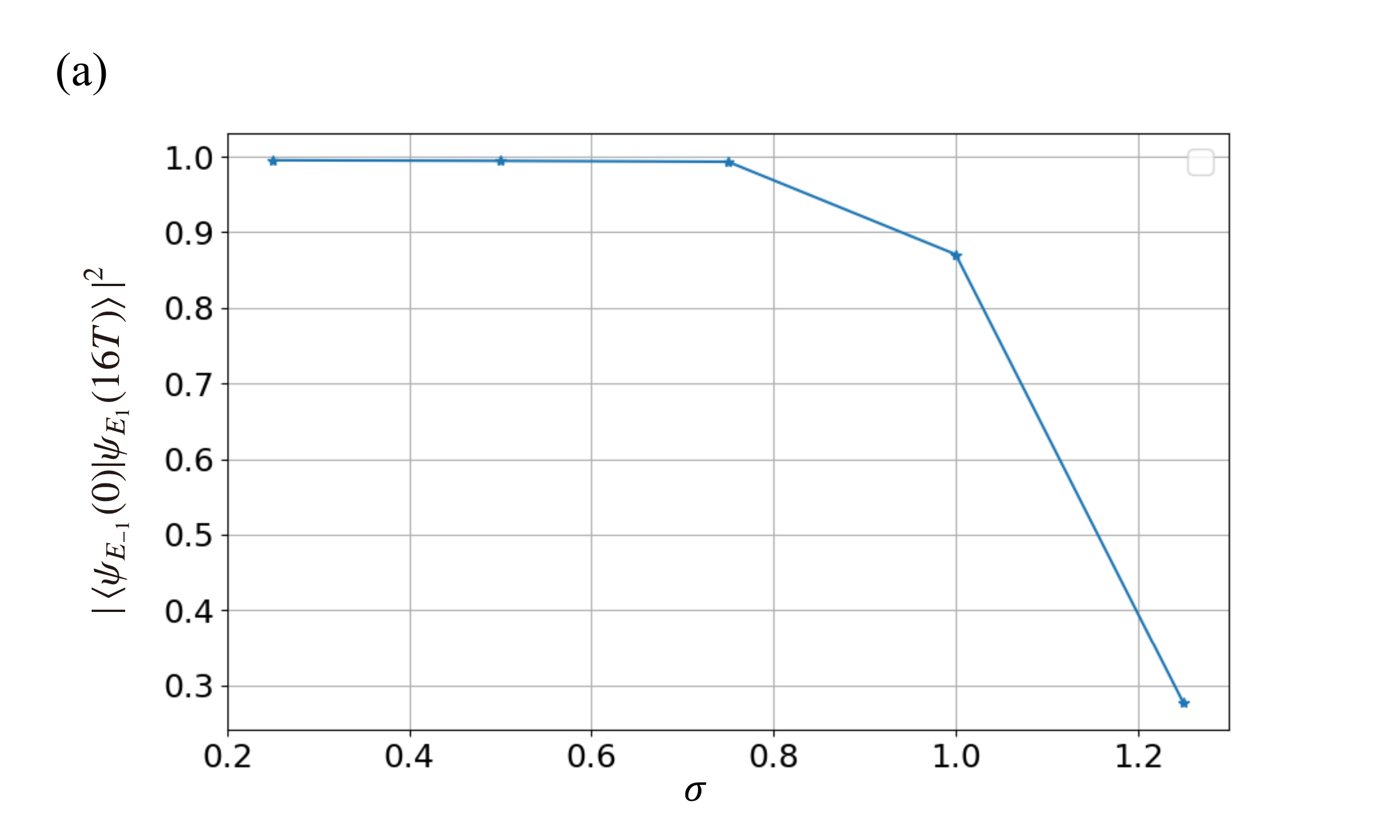}
    \includegraphics[width=90mm]{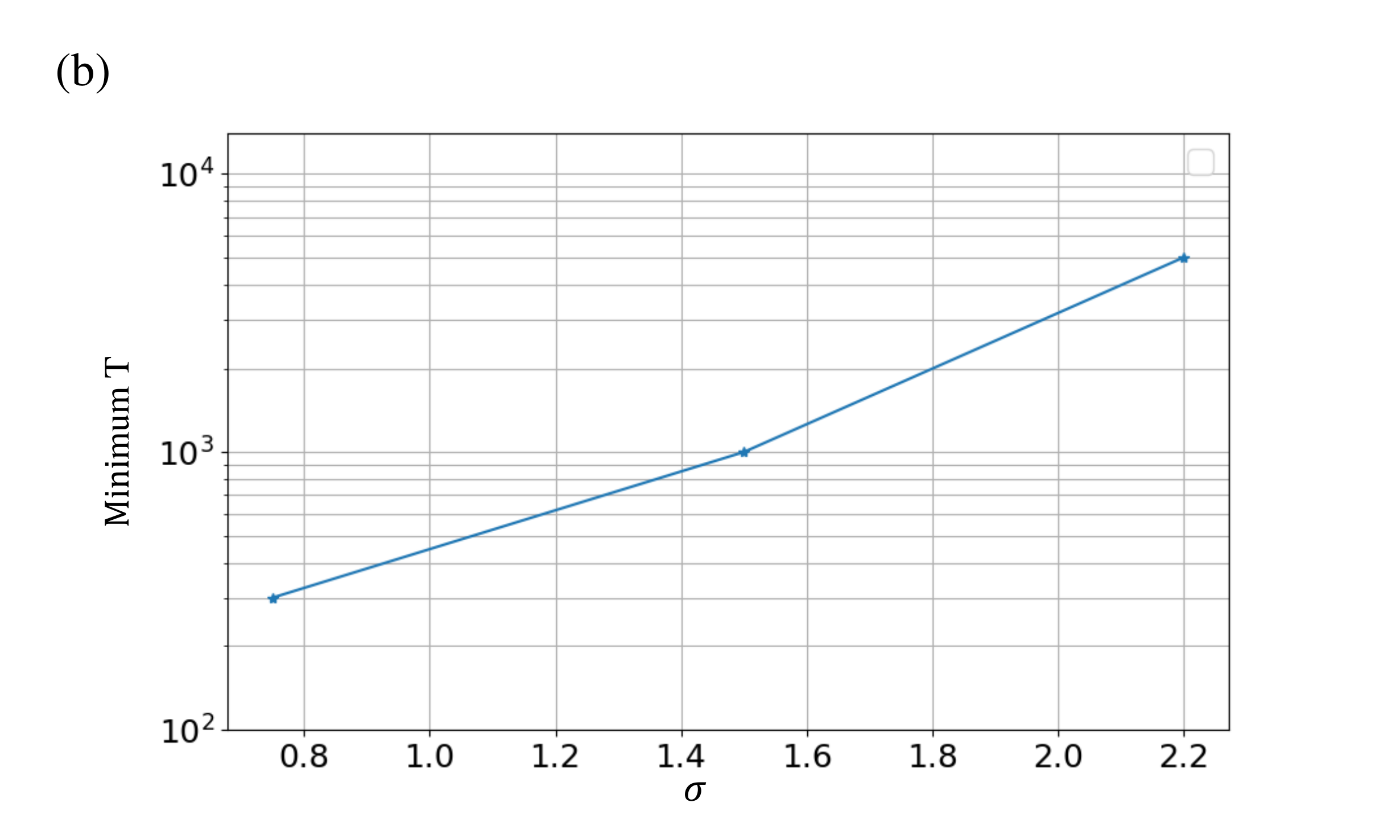}
    \caption{(a) Transition probability between the $E_1$ state and the $E_{-1}$ state,  
    which measures the probability of achieving the Majorana NOT gate, plotted as a function of $\sigma$. (b) Minimum $T$ required for the transition probability $\sim 1$, corresponding to the non-Abelian braiding. We take system parameters for (a) and (b) as $t_h=1$, $\mu_{\mathrm{min}}=0.1$, $\mu_{\mathrm{max}}=7$, $\alpha_R=5$, $\Delta_0=10$, $L_{\mathrm{wire}}=40$ and $V_0=5\mu_{\mathrm{max}}$.}
    \label{fig:gate_induced}
\end{figure}



\begin{figure*}[t!]
    \includegraphics[width=150mm]{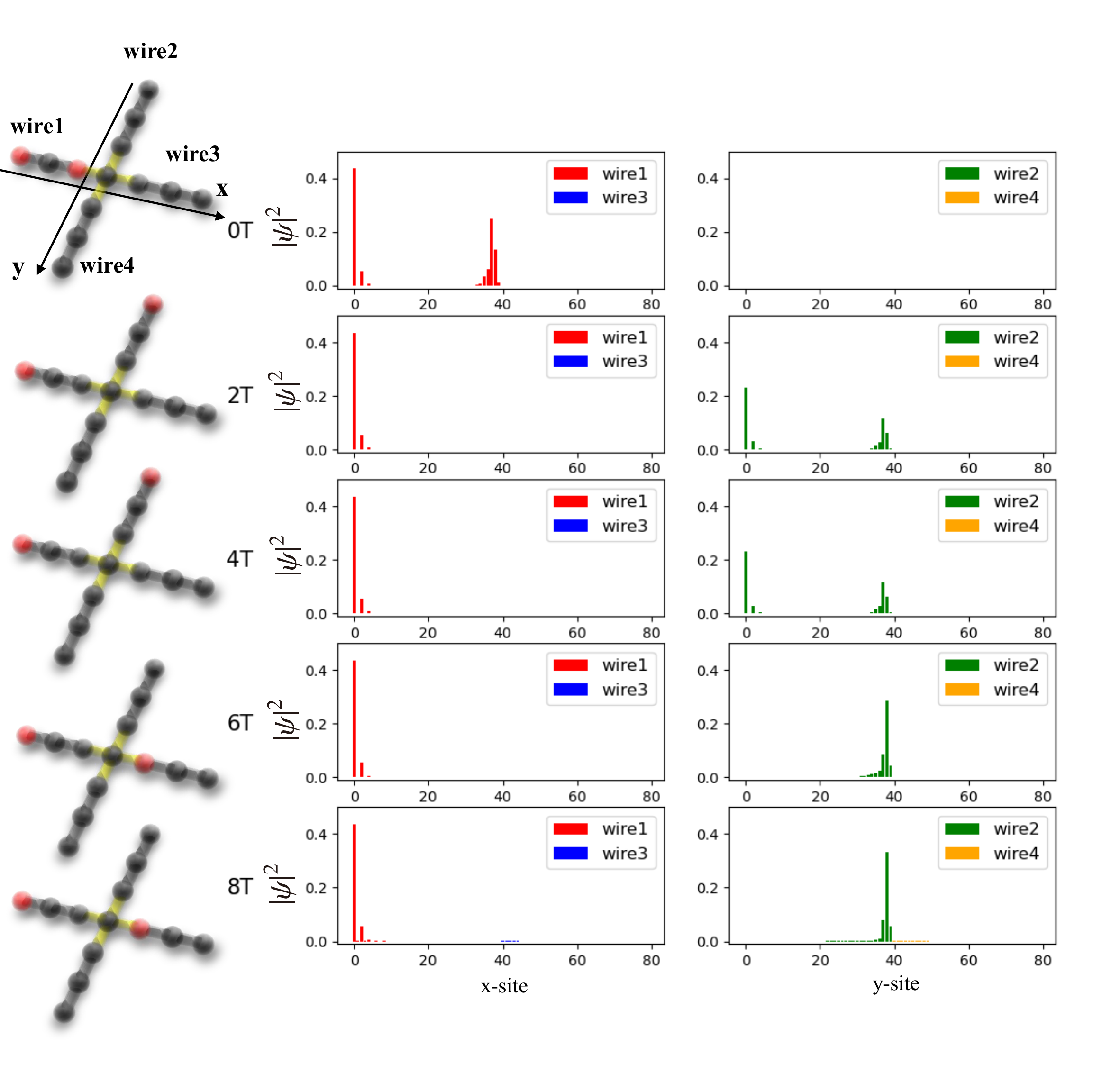}
    \caption{ Time evolution of the profile of the wave function of low-energy bound states 
    for the failed braiding process in the case with gate-induced inhomogeneous potentials.
    We set the width of the potential as $\sigma=1.25$. In this case, non-Majorana low-energy Andreev bound states exist near the gates. 
    The braiding process of the state belonging to one Kramers pair partner  in the wire 1 is shown. 
    The other parameters are the same as (a). 
    A MKP "sticks" in the vicinity of one end of the wire 2 for the intermediate steps from $t\sim 0$ to $t=8T$.}
    \label{fig:braiding_process_gate_induced2}
\end{figure*}


From the simulations of the braiding dynamics in the case with the gate-induced inhomogeneous potential, it is found that the existence of gate-induced potential can disturb the transfer of Majorana quasi-particles and that the characteristic length controlling whether the braiding process is successful or not is the Majorana coherence length $\xi$, which is nearly equal to the coherence length of the superconductor. We plot the transition probabilities $\abs{ \braket{\psi_{E_1}(0)}{\psi_{E_{-1}}(16T)}}^2$  for the NOT gate
as a function of $\sigma$ in Fig.~\ref{fig:gate_induced}(a). 
The data show that the probabilities start to deviate from the ideal value corresponding 
to the successful braiding $\abs{ \braket{\psi_{E_1}(0)}{\psi_{E_{-1}}(16T)} }^2\sim 1$,
 when $\sigma$ exceeds
the Majorana coherence length $\xi \sim 1$, 
which is estimated from the profile of the wave function of Majorana quasi-particle shown in Fig.~\ref{fig:braiding_process_success}. 

In Fig.~\ref{fig:braiding_process_gate_induced2}, we show the time evolution of the profile of the wave function of low-energy states for the failed braiding process. In the case with gate-induced potentials, non-Majorana low-energy Andreev bound states exist near the gates of the wire 2 and wire 4, which are in the trivial phase in the initial state, while the wire 1 and wire 3 are in the topological phase, hosting MKPs. The non-Majorana low-energy states can have non-trivial influence on the transfer of MKPs from the wire 1 to wire 2. In fact, it can be seen through our simulations that a MKP "sticks" near an end of the wire 2, where non-Majorana low-energy Andreev bound states exist, in the intermediate steps from $t\sim 2T$ to $t \sim 8T$, which leads to the disturbance of the braiding process. This behavior indiates that quasi-particle poisoning due to the existence of low-energy non-Majorana states cannot be ignored in this case. This is contrasted with successful braiding process in the absence of low-energy bound states, as shown in Fig.~\ref{fig:braiding_process_success}. 
In addition, we calculate the dependence of required $T$ for sustain transition probability on $\sigma$, which is shown in Fig.~\ref{fig:gate_induced}(b). 
The result implies that required $T$ non-linearly increases as $\sigma$ does. 
However, the influence can be ignored when the width of the gate-induced potential $\sigma$ is sufficiently small compared to the Majorana coherence length $\xi\sim1$, as shown in
Fig.~\ref{fig:gate_induced}(a).
Thus, for the realization of the successful braiding, it is crucially important to control the length scale of the gate-induced potentials of junction systems.
We would like to note that our results are not qualitatively changed even if we use other forms of an inhomogeneous potential, provided that the potential has only one characteristic length scale, which determines the localization length of the non-Majorana Andreev bound states.

\section{CONCLUSION}
By performing numerical dynamical simulations of braiding processes, we have investigated the tolerance of the non-Abelian statistics and the Majorana quantum gates against two types of perturbations which may give rise to decoherece of  MKP qubits in time-reversal invariant topological superconductor systems.
One perturbation is a magnetic field which breaks time-reversal symmetry, and the other one is a gate-induced inhomogeneous potential which induces
non-Majorana low-energy Andreev bound states at junctions of the superconducting nanowire system.
We have revealed the following three points. (i) MKPs obey non-abelian statistics in ideal situations. (ii) Even in the case with an applied magnetic field, when the field direction preserves the combination of time-reversal symmetry and mirror symmetry in the initial and final states of a braiding process, 
the non-Abelian braiding is successful.
Remarkably, this tolerance is preserved even when the combined symmetry is broken in the intermediate states of the braiding process.
(iii) Non-Majorana low-energy Andreev bound states generated by gate-induced inhomogeneous potentials interfere with MKPs and prevent MKPs from moving properly in the braiding process, which makes Majorana qubits vulnerable to quasi-particle poisoning and disturbs the braiding protocol. However, the influence can be ignored when the width of the gate-induced potential is sufficiently smaller than the Majorana coherence length, which is roughly given by
the coherence length of the superconductor.

Although we concentrated on Majorana particles in a TRITSC in this paper, 
it is expected that effects of non-Majorana low-energy states generated by inhomogeneous gate potentials on the non-Abelian braiding 
is ubiquitous for any topological superconductors including nanowire systems under applied magnetic fields
~\cite{Mourik_2012,Deng_2012,das12,PhysRevLett.119.136803,gul,deng16,finck,churchill}.
Thus, it is important for the construction of Majorana quantum gates to control properly the length scale of gate-induced potentials at junctions
of nanowire systems.

\begin{acknowledgments} 
A part of the computation in this work has been done by using the facilities of the Supercomputer Center, the institute for Solid Physics, the University of Tokyo. T.S. was supported by a JSPS Fellowship for Young Scientists. This work was supported by JST CREST Grant No. JPMJCR19T5, Japan, and the Grant-in-Aid for Scientific Research on Innovative Areas ``Quantum Liquid Crystals (No.~JP20H05163 and No.~JP22H04480)'' from JSPS of Japan, and JSPS KAKENHI (Grants No.~JP20K03860, No.~JP20H01857, No.~JP21H01039, and No.~JP22H01221).

\end{acknowledgments}

\bibliography{classDIII_braiding.bib}
\end{document}